\documentclass[a4paper,12pt]{article}

\usepackage{a4}
\usepackage{graphics}
\usepackage{epsfig}
\usepackage{amssymb}

\begin{document}
\hfill WUE-ITP-2003-003\\
\hspace*{\fill} IPPP-03-03\\
\hspace*{\fill} DCPT-03-06\\
\hspace*{\fill} hep-ph/0303044\\
\begin{center}
{\Large \bf Precision measurements of
Higgs-chargino couplings in chargino pair production at a muon collider}
\end{center}
\vspace{0.5cm}
\begin{center}
{\large \sf H. Fraas$^*$, F. Franke$^*$,
G. Moortgat-Pick$^{\dagger}$, F. von der Pahlen$^*$, A.~Wagner$^*$}
\end{center}
\begin{center}
{\small \it $^*$ Institut f\"ur Theoretische Physik, Universit\"at W\"urzburg, Am 
Hubland, \\
D-97074 W\"urzburg, Germany\\
$^{\dagger}$ Institute for Particle Physics and Phenomenology, University of Durham, \\
Durham DH1 3LE, UK}

\end{center}
\begin{abstract}
We study chargino pair production on the heavy Higgs resonances at
a muon collider in the MSSM. At $\sqrt{s} \approx 350$~GeV cross
sections up to 2 pb are reached depending on the supersymmetric
scenario and the beam energy spread. The resonances of the scalar
and pseudoscalar Higgs bosons may be separated for $\tan\beta <8$.
Our aim is to determine the ratio of the chargino couplings to the
heavy scalar and pseudoscalar Higgs boson independently of the
specific chargino decay characteristics. The precision of the
measurement depends on the energy resolution of the muon collider
and on the error in the measurement of the cross sections of the
non-Higgs channels including an irreducible standard model background. 
With a high energy resolution the systematic
error can be reduced to the order of a few percent.
\end{abstract}

\section{Introduction}

Since a muon collider produces Higgs bosons directly via $\mu^+\mu^-$
annihilation in the s-channel, it is an excellent tool to study the
properties of a heavy scalar or pseudoscalar Higgs boson
\cite{hefreport1,hefreport2,mucolhiggs,barger1}.
Especially the determination of the Higgs couplings constitutes an
important test of supersymmetric models. In this paper we explore the
potential of a muon collider for a precision measurement of the
Higgs-chargino couplings in the Minimal Supersymmetric Standard Model
(MSSM). Therefore we focus on the chargino pair production in order to
determine the chargino couplings to the exchanged Higgs boson in the s-channel.

In the MSSM charginos are the mass eigenstates formed by the mixing
of the supersymmetric partners of the charged $W$ and Higgs bosons. 
While their masses and mixing can be determined with high precision
at an $e^+e^-$ collider \cite{gudi choi kneur,lc}, 
a muon collider is the by far more suitable 
machine to study their couplings to Higgs bosons. 
The MSSM contains three neutral Higgs bosons: a light scalar $h$ and two
heavier Higgs particles, a scalar $H$ and a pseudoscalar $A$.  Higgs bosons
are expected to be discovered at the LHC and studied in the clean
environment of a linear $e^+e^-$ collider. However, a linear collider will
probably not reveal all properties of the heavy supersymmetric Higgs bosons
in detail.
The cross sections for the processes $e^+e^- \rightarrow Z\{H,A\}$ are
heavily suppressed close to the Higgs decoupling limit \cite{decoupling}.  
The main production mechanism for heavy Higgs bosons is the associated
production $e^+e^- \rightarrow HA$, which yields cross sections in the fb
range \cite{lc}.
But for a subsequent determination of the Higgs chargino couplings one has
to discriminate between charginos from $H$ and $A$ decay.  Here it turns
out to be rather complicated to find observables which allow to identify
the CP quantum number of the mother particle of the chargino pairs
\cite{djouadi}.
For beam energies below the $HA$ threshold single Higgs production 
$ e^+e^- \rightarrow H\nu\bar\nu$ has been studied in \cite{singlehiggsh}.
The small cross section of this process, however, significantly restricts
the potential for precision studies of the Higgs properties.

Also the $\gamma\gamma$ mode of a linear collider will not be suitable for
a precise measurement of the heavy Higgs bosons to charginos. Although $H$
or $A$ can be resonantly produced, the background from chargino pair
production $\gamma\gamma \rightarrow \tilde{\chi}^+\tilde{\chi}^-$ is one
order of magnitude larger than the signal $\gamma\gamma \rightarrow H,A
\rightarrow \tilde{\chi}^+\tilde{\chi}^-$ \cite{gammagamma}.
Furthermore one has to deal with a significantly larger energy spread
compared to a muon collider.

A muon collider could overcome these difficulties 
by providing a heavy Higgs factory \cite{hefreport1,hefreport2}. 
In a relevant part of the parameter space the Higgs branching ratios for
the decay into chargino pairs is sufficient to perform precise measurements
of the Higgs chargino couplings.  In order to be independent of the
specific chargino decay mechanism we focus on the ratio of the chargino
couplings to the scalar and pseudoscalar Higgs.
Then the relevant observables are merely the total cross sections at the
$H$ and $A$ resonances and the contribution of the non-Higgs channels that
can be measured without any model-dependent assumptions.

The achievable precision is generally limited by the energy resolution
of the muon collider and the separation of the relevant Higgs channel
from the non-resonant contribution in the chargino production process.
An essential requirement is that the $H$ and $A$ signals can be
clearly separated. Therefore we also study the overlap of the Higgs
resonances as a function of the energy resolution and $\tan\beta$.

This paper is organized as follows: In Section 2 we give analytical
formulae for the cross sections and characterize the observables for
the determination of the Higgs-chargino couplings.  Section 3 contains
numerical results for representative supersymmetric scenarios with
different chargino mixing, Higgs masses and values of $\tan\beta$.  We
show cross sections for the pair production of the light chargino
$\mu^+\mu^- \rightarrow \tilde{\chi}_1^+\tilde{\chi}_1^-$ and estimate
the relative systematic error in the determination of the Higgs
chargino couplings.

\section{Analytical formulae}

\subsection{Lagrangians and cross sections}

We study chargino  pair production in the MSSM 
\begin{equation}\label{prod}
\mu^+\ \mu^-\rightarrow\tilde{\chi}_i^+\ \tilde{\chi}_j^-
\end{equation}
for CMS-energies $\sqrt{s}$ at the resonances of the heavy neutral Higgs
bosons $H$ and $A$.

This process proceeds via the exchange of $H$ and $A$ in the s-channel,
whereas the contribution from the exchange of the gauge bosons $\gamma$,
$Z$ and of the light Higgs boson $h$ in the s-channel as well as from the 
t-channel exchange of $\tilde{\nu}_\mu$ constitutes the background in our
analysis.

The interaction Lagrangians for chargino production via Higgs exchange 
are
\begin{eqnarray}
\label{eq mumuphi}
{\cal L}_{\mu^+ \mu^- \phi} & = &
          g \ c^{(\phi\mu)}
			\bar{\mu} \ \Gamma^{(\phi)} \ \mu \ \phi,\\
{\cal L}_{\tilde{\chi}^{\pm} \tilde{\chi}^{\pm} \phi} & = &
          g \ \bar{\tilde{\chi}}_i^+  
          (c^{(\phi)L}_{ij} P_L + c^{(\phi)R}_{ij} P_R) 
           \tilde{\chi}_j^+\phi
\label{eq chichiphi}
\end{eqnarray}
with $\phi=H,A,h$,
$\Gamma^{(H)}=\Gamma^{(h)}=1$, $\Gamma^{(A)}=i\gamma^5$ 
and implicit summation over $i,j$. 

Explicit expressions for the Higgs-muon couplings $c^{(\phi\mu)}$ can
be found in \cite{GH}.  They are determined by the Higgs mixing angle
$\alpha$ and by the ratio of the vacuum expectation values of the two
neutral Higgs fields $\tan\beta=v_2/v_1$. The Higgs-chargino couplings
$c^{(\phi)L}_{ij}= c^{(\phi)R\ast}_{ji}$ \cite{GH} depend on
$\tan\beta$, the SU(2) gaugino mass $M_2$ and the higgsino mass
parameter $\mu$ that determine the masses and the mixing characters of
the charginos.

In the MSSM with CP-conservation
the interference between 
$H$ and $A$ exchange vanishes.
Furthermore the interference between the Higgs boson exchange
and the $\gamma$, $Z$ and $\tilde{\nu}_{\mu}$ channels
is strongly suppressed by a factor
$m_{\mu}/\sqrt{s}$. Therefore the total cross section of the
production of chargino pairs 
$\tilde{\chi}^+_i \tilde{\chi}^-_j$ 
can be separated into the dominating contributions
$\sigma_H^{ij}$ and $\sigma_A^{ij}$ from $H$ and $A$ exchange
and the background $\sigma^{ij}_{B,SUSY}$ from 
$\gamma$, $Z$, $\tilde{\nu}_\mu$ and $h$ exchange
\begin{equation}
\sigma^{ij} = \sigma_{H}^{ij} +  \sigma_{A}^{ij} + \sigma_{B,SUSY}^{ij}.
\label{sigmasep}
\end{equation}

Chargino production via the $\gamma$, $Z$, $\tilde{\nu}_\mu$ channels will have been thoroughly studied at linear colliders \cite{lc}.  The $h$ exchange contribution can be neglected at the $H$ and $A$ resonances.
 
At CMS energy $\sqrt{s}$ the cross sections
$\sigma_H^{ij}$ and $\sigma_A^{ij}$ are
\begin{eqnarray}
\sigma_{\phi}^{ij} & = &  \frac{g^2}{4\pi}\vert c^{(\phi\mu)}\vert^2\cdot
\vert c^{(\phi)R}_{ij}\vert^2\cdot B_{\phi}^{i,j}(s)K_{\phi}(s),\quad \phi=H,A
\label{siphi}
\end{eqnarray}
with
\begin{eqnarray}
          K_{\phi}(s) &=&
     \frac{s}{(s-m_\phi^2)^2 + \Gamma_\phi^2 m_\phi^2},
\label{propphi}
\\
	B_{H}^{ij}(s) &=& \frac{\lambda(s,m_i^2,m_j^2)^{3/2}}{s^3},	
\\
	B_{A}^{ij}(s) &=& \frac{\lambda(s,m_i^2,m_j^2)^{1/2}}{s},		
\\
\lambda(s,m_i^2,m_j^2)&=& s^2-2s(m_i^2+m_j^2)-(m_i^2-m_j^2)^2
\label{lambda}
\end{eqnarray}

The total cross section $\sigma^{f_+f_-}$ for the pair production 
$\mu^+\mu^-\rightarrow\tilde{\chi}_i^+\tilde{\chi}_j^-$
with subsequent decays $\tilde{\chi}_i^+\rightarrow f_+$ 
and $\tilde{\chi}_j^-\rightarrow f_-$  factorizes
into the production cross section $\sigma^{ij}$ and the 
branching ratios for the respective decay channels: 
\begin{equation}
 \sigma^{f_+f_-}(\sqrt{s}) = \sigma^{ij}(\sqrt{s})\times BR(\tilde{\chi}_i^+\rightarrow f_+)\times BR(\tilde{\chi}_j^-\rightarrow f_-).
\end{equation}
This holds for each of the contributions 
$\sigma_H^{f_+f_-}$ from $H$ exchange, 
$\sigma_A^{f_+f_-}$ from $A$ exchange
and $\sigma_{B,SUSY}^{f_+f_-}$ from the background channels in eq.~(\ref{sigmasep}).

\subsection{Determination of the Higgs-chargino couplings}

In the following we consider the pair production of the light chargino
$\tilde{\chi}_1^{\pm}$ that is expected to be among the first
kinematically accessible supersymmetric particles at a muon collider.
In order to determine the Higgs-chargino couplings one has to separate
the Higgs exchange contributions $\sigma_H^{f_+f_-}+\sigma_A^{f_+f_-}$
from the total measured cross sections $\sigma_{meas}^{f_+f_-}$, at $\sqrt{s}=m_H$ and
$\sqrt{s}=m_A$, respectively.  Since the interference between the
Higgs channels and the background is negligible we can subtract the
contributions $\sigma_{B,SUSY}^{f_+f_-}$ 
from the total cross section.

Besides the non-resonant contributions to the chargino pair
production one has to consider further background sources
from standard model processes.
Here $W$ pair production and single $W$ production constitute the main standard model
background, which is in principle rather large \cite{lc} but
can be strongly reduced by appropriate cuts
\cite{smbackground}.  Then the resonance
peaks remain clearly visible above the smooth standard model background $\sigma_{B,SM}^{f_+f_-}$ 
which can therefore be included in the
subtraction of the non-resonant contribution from the total
cross section.  

We determine the total background contribution  $\sigma_{B}^{f_+f_-}=\sigma_{B,SUSY}^{f_+f_-}+\sigma_{B,SM}^{f_+f_-}$
by linear interpolation of $\sigma_{meas}^{f_+f_-}$ far below and above the
resonance energies. The precision of this estimate obviously depends
on the variation of the background contributions around the heavy
Higgs resonances.  By this procedure we avoid, however, reference to
other experiments at different energy scales as e.~g.~chargino
production at $e^+e^-$ colliders combined with specific model
calculations.

Due to their factorization into production and decay
the ratio of the measured contribution from $H$ and $A$ exchange 
\begin{equation}
r=\frac{\sigma_{meas}^{f_+f_-}(m_H) - \sigma^{f_+f_-}_{B}(m_H)}{\sigma_{meas}^{f_+f_-}(m_A) - \sigma^{f_+f_-}_{B}(m_A)}=\frac{\sigma^{11}_H(m_H) + \sigma^{11}_{A}(m_H)}{\sigma^{11}_H(m_A) + \sigma^{11}_{A}(m_A)}
\label{rratio}\end{equation} 
is independent of the specific chargino decay channel which may be
chosen to give the best experimental signal.
Then the measurement of the total cross section for chargino production 
and decay at the Higgs resonances offers
an interesting possibility to determine the ratio of the
Higgs-chargino couplings
\begin{equation}
x=\left(\frac{c^{(H)R}_{11}}
{c^{(A)R}_{11}}\right)^2.
\label{xratio}
\end{equation}

From eqs.~(\ref{siphi}) and (\ref{rratio}) one obtains
\begin{equation}
x=\frac{r}{C}\cdot\frac{1-C_1/r}{1-C_2/r}\cdot\frac{1}{x_\mu},
\end{equation}
with
\begin{eqnarray}
C & = & \frac{\beta^3(m_H^2)}{\beta(m_A^2)}\frac{\Gamma_A^2}{\Gamma_H^2},\\
C_1 & = & \frac{\beta(m_H^2)}{\beta(m_A^2)}K_A(m_H^2)\Gamma_A^2,\\
C_2 & = & \left(\frac{\beta(m_A^2)}{\beta(m_H^2)}\right)^3K_H(m_A^2)\Gamma_H^2,\\
\beta(s)&=& \left(\frac{\lambda(s,m_1^2,m_1^2)}{s^2}\right)^{1/2}
	 = \left( \frac{s - 4 m_{\tilde{\chi}_1^{\pm}}^2}{s} \right)^{1/2},
\label{beta}\\
x_\mu&= &\left(\frac{c^{(H\mu)}}{c^{(A\mu)}}\right)^2,
\label{xmuratio}
\end{eqnarray}
where $C,\ C_1$ and $C_2$ can be determined without model 
dependent assumptions,
and $x_\mu=1$ 
in the Higgs decoupling limit.

Assuming that the masses of the heavy Higgs bosons and the chargino are precisely known \cite{lc,barger2} 
the precision for the determination of $x$ depends on the energy spread
of the muon beams, the width of the $H$ and $A$ resonances 
and on the error in the determination of the background. 

\section{Numerical results}
In the numerical analysis we estimate how precisely the ratio
of the couplings of the lighter chargino to the heavy Higgs bosons $H$ and $A$ can be measured. 
We study the cross sections for 
the production $\mu^+\mu^-\rightarrow\tilde{\chi}_1^+\tilde{\chi}_1^-$ 
of the lighter
chargino with unpolarized beams.

The mass of the scalar Higgs bosons,
the widths of $A$ and $H$ and the branching ratios for their
decays into charginos are computed with the program HDECAY \cite{HDECAY}.
The matrix elements of the unitary $2\times 2$ matrices
that diagonalize the chargino mass matrix are defined by
$U_{ij}$ and $V_{ij}$ \cite{HK}.

\subsection{Scenarios}

We choose 
six representative scenarios A -- F
with $m_{\tilde{\chi}_1^{\pm}}=155$~GeV, 
$m_A=350$~GeV, and $\tan\beta=5$ 
which differ by the mixing characteristic of the chargino and by
the sign of the higgsino mass parameter $\mu$.
The parameters, masses 
and the gaugino and higgsino contents of $\tilde{\chi}_1^{\pm}$ are given
in table \ref{szentab1}.
In scenarios A with $\mu<0$ and B with $\mu>0$ 
the light chargino is a wino-higgsino
mixing. In scenarios C ($\mu<0$) and D ($\mu>0$)
it has a dominant gaugino character whereas in scenarios
E ($\mu<0$) and F ($\mu>0$) it is nearly a pure higgsino.

The additional scenarios in table \ref{szentab2} are derived from
the mixed scenario B and the gaugino scenario C by varying
$\tan\beta$ and the masses of the light chargino and the
pseudoscalar Higgs boson.
\begin{table}
\renewcommand{\arraystretch}{1.2}
\begin{center}
\begin{tabular}{|c||c|c|c|c|c|c|}
\hline
{\bf Scenarios} & {\bf A}& {\bf B}& {\bf C}& {\bf D}& {\bf E}& {\bf
F}\\
\hline
\hline
$M_2$/GeV & 188 & 217.3 & 154.9 & 169.5 & 400 & 400 \\
\hline
$\mu$/GeV & -188 & 217.3 & -400 & 400 & -154.9 & 169.5 \\
\hline
$U_{11}$ & 0.577 & -0.632 & 0.958 & -0.943 & 0.056 & -0.184\\
\hline
$U_{12}$ & 0.817 & 0.775 & 0.288 & 0.333 & 0.9984 & 0.983\\
\hline
$V_{11}$ & 0.817 & 0.775 & 0.9984 & 0.983 & 0.288 & 0.333\\
\hline
$V_{12}$ & -0.577 & -0.632 & -0.056 & -0.184 & -0.958 & -0.943\\
\hline
\hline
$m_H$/GeV & 352.1 & 352.3 & 351.9 & 352.3 & 352.2 & 352.3\\
\hline¬
$\Gamma_H$/GeV & 0.67 & 0.58 & 0.31 & 0.32 & 0.32 & 0.39\\
\hline¬
$\Gamma_A$/GeV & 1.05 & 1.33 & 0.43 & 0.57 & 0.43 & 0.64\\
\hline¬
$c_{11}^{(H)R}$ & 0.513 & 0.347 & 0.207 & 0.197 & 0.207 & 0.197 \\
\hline¬
$c_{11}^{(A)R}$ & 0.417 & 0.472 & 0.192 & 0.251 & 0.192 & 0.251 \\
\hline
\end{tabular}
\end{center}
\caption{
Reference scenarios with fixed
$m_A=350$~GeV, $m_{\tilde{\chi}_1^{\pm}}=155$~GeV, $\tan\beta=5$ and
$m_{\tilde{\nu}_\mu}=261.3$~GeV.
$U_{11}$ and $V_{11}$ ($U_{12}$ and $V_{12}$) are the gaugino
(higgsino) components·
of the charginos \cite{HK}. 
$c_{11}^{(H)R}$ and $c_{11}^{(A)R}$ denote the Higgs-chargino couplings.
}
\renewcommand{\arraystretch}{1.0}
\label{szentab1}
\end{table}
\begin{table}
\centering
\renewcommand{\arraystretch}{1.2}
\begin{tabular}{|c||c|c|c|c|c|c|c|c|}
\hline
{\bf Scenarios} & {\bf B400}& {\bf C400}& {\bf B180}& {\bf C180}& {\bf B7}&
{\bf B8}& {\bf C7}& {\bf C8}\\
\hline
\hline
$M_2$/GeV & 217.3 & 154.9 & 242.8 & 180.7 & 214 & 212.8 & 156.9 &
157.5 \\
\hline
$\mu$/GeV & 217.3 & -400 & 242.8 & -420 & 214 & 212.8 & -400 & -400
\\
\hline
$\tan\beta$ & 5 & 5 & 5 & 5 & 7 & 8 & 7 & 8 \\
\hline
$m_{\tilde{\chi}_1^{\pm}}$ & 155 & 155 & 180 & 180 & 155 & 155 & 155 &
155 \\
\hline
$U_{11}$  & -0.632 & 0.958 & -0.640 & 0.959  & -0.625 & -0.622 & 0.955
& 0.954\\
\hline
$U_{12}$  & 0.775  & 0.288 & 0.768  & 0.283  & 0.781  & 0.783  & 0.297
& 0.300\\
\hline
$V_{11}$  & 0.775  & 0.9984 & 0.768  & 0.9977  & 0.781  & 0.783  &
0.9972 & 0.9967\\
\hline
$V_{12}$  & -0.632 & -0.056 & -0.640 & -0.068 & -0.625 & -0.622 &
-0.075 & -0.081\\
\hline
\hline
$m_A$/GeV & 400 & 400 & 400 & 400 & 350 & 350 & 350 & 350\\ 
\hline
$m_H$/GeV & 402.0 & 401.6 & 402.0 & 401.6 & 351.2 & 350.9 & 351.0 &
350.7\\
\hline
$\Gamma_H$/GeV & 1.17 & 0.61 & 0.82 & 0.52 & 0.71 & 0.80 & 0.44 &
0.53\\
\hline
$\Gamma_A$/GeV & 2.43 & 1.09 & 1.96 & 1.00 & 1.42 & 1.50 & 0.57 &
0.67\\
\hline
\end{tabular}
\caption{
Reference scenarios with different mass $m_A$ (scenarios
B400 and C400),
with different masses $m_{\tilde{\chi}_1^{\pm}}$ and $m_A$, (scenarios
B180 and C180)
and different values of $\tan\beta$ (scenarios
B7, B8 and  C7, C8)
as in the reference scenarios (table \ref{szentab1}).
$U_{11}$ and $V_{11}$ ($U_{12}$ and $V_{12}$) are the gaugino
(higgsino) components·
of the charginos \cite{HK}.}
\renewcommand{\arraystretch}{1.0}
\label{szentab2}
\end{table}

In order to study the influence of the Higgs mass, $m_A$ is increased
from $m_A=350$~GeV to $m_A=400$~GeV in scenarios B400 and C400.
The influence of the chargino mass will be analyzed with the help
of scenarios B180 and C180 where 
$m_{\tilde{\chi}_1^\pm}=180$~GeV and $m_A=400$~GeV in order to ensure
$m_A>m_{\tilde{\chi}_1^\pm}/2$. However, the character of
the light chargino is nearly identical in scenarios B, B180 and B400 (gaugino-higgsino mixing)
and in scenarios C, C180 and C400 (gaugino like), respectively.

Finally we study the influence of higher values of
$\tan\beta=7$ and $\tan\beta=8$ for
$m_A=350$~GeV and $m_{\tilde{\chi}_1^\pm}=155$~GeV in scenarios
B7, B8 and C7, C8. 
To obtain  a similar
chargino mixing character the parameters $M_2$ and $\mu$ are slightly changed compared
to scenarios B and C with
$\tan\beta=5$. 

\subsection{Branching ratios and cross sections}

\begin{figure}[ht]
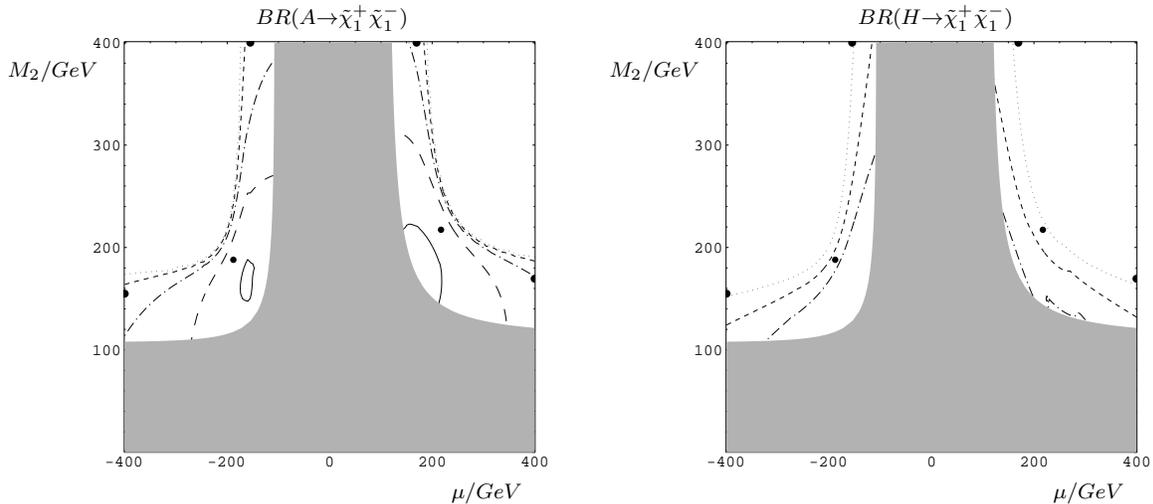

\centering
\setlength{\unitlength}{1cm}
\begin{picture}(12,6.3)
\put(-.5,0.){\includegraphics{fig.1a}}
\put(7.5,0.){\includegraphics{fig.1b}}
\put(1.7,6.1){$ \scriptstyle BR(A \rightarrow \tilde{\chi}^+_1 \tilde{\chi}^-_1)$}
\put(-1.6,5.4){$ \scriptstyle M_2/GeV$}
\put(4.3,-.2){$ \scriptstyle \mu/GeV$}
\put(9.7,6.1){$ \scriptstyle BR(H \rightarrow \tilde{\chi}^+_1 \tilde{\chi}^-_1)$}
\put(6.4,5.4){$ \scriptstyle M_2/GeV$}
\put(12.3,-.2){$ \scriptstyle \mu/GeV$}
\end{picture}
\caption{\small
        Branching ratios of the heavy Higgs bosons
        $A$ and $H$
        into light chargino pairs for $m_A=350$~GeV, $\tan\beta=5$ and sfermions masses larger than $ M_H/2 $,
        computed with the program HDECAY \cite{HDECAY}.
        The contour lines correspond to
        0.1 (dotted), 0.2 (dashed), 0.3 (dash-dotted),
        0.4 (large dashed) and 0.5 (solid).
        The gray area is the experimentally excluded region
	given here by $m_{\tilde{\chi}_1^\pm}<100 \mbox{ GeV}$,
        the thick dots are the scenarios {A}~--~{F} 
	of table \ref{szentab1}. 
	}
\label{br_contour}
\end{figure}

The branching ratios for the decays of the Higgs bosons $H$ and $A$
into a light chargino pair are crucial for obtaining sufficient cross sections.
Therefore we show in fig.~\ref{br_contour} contour plots for the branching ratios in
the $M_2-\mu$ plane for $\tan\beta=5$ and $m_A=350$~GeV and indicate
our scenarios A -- F.

Since the Higgs bosons couple to both the gaugino and higgsino component
of the chargino, 
the couplings and branching ratios are large in the		
parameter region $|M_2| \approx |\mu|$ of the mixed scenarios A and B.
In scenario A (B) with $\mu < 0$ ($\mu>0$) one obtains branching
ratios up to 45\% (20\%) for the $A$ decay and up to 
20\% (15\%) for the $H$ decay.
In scenarios C and D with a gaugino dominated light chargino
as well as in scenarios E and F with a higgsino-like light chargino
branching ratios between 20\% and 30\% for the $A$ decay and between
10\% and 20\% for the $H$ decay
can be observed.

\begin{figure}[ht]
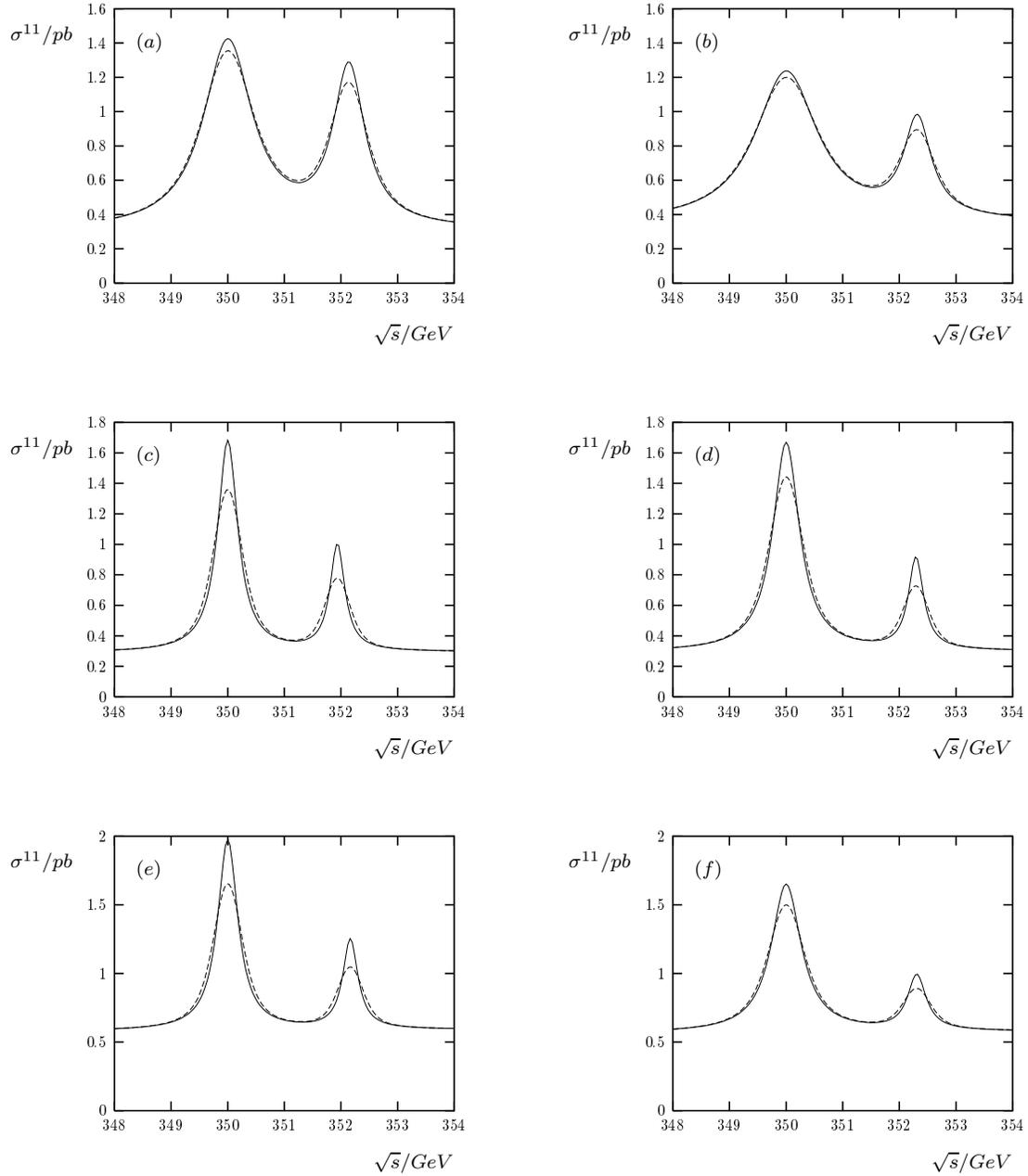

\centering
\setlength{\unitlength}{1cm}
\begin{picture}(12,5.9)
\put(-2.5,-9.05){\includegraphics{fig.2a}}
\put(5.5,-9.05){\includegraphics{fig.2b}}
\put(-1.3,4.3){$\scriptstyle \sigma^{11}/pb$}
\put(3.9,0.){$ \scriptstyle \sqrt{s}/GeV$}
\put(6.7,4.3){$\scriptstyle \sigma^{11}/pb$}
\put(11.9,0.){$ \scriptstyle \sqrt{s}/GeV$}
\put(.5,4.2){$ \scriptstyle (a)$}
\put(8.5,4.2){$ \scriptstyle (b)$}
\end{picture}
\begin{picture}(12,5.9)
\put(-2.5,-9.05){\includegraphics{fig.2c}}
\put(5.5,-9.05){\includegraphics{fig.2d}}
\put(-1.3,4.3){$\scriptstyle \sigma^{11}/pb$}
\put(3.9,0.){$ \scriptstyle \sqrt{s}/GeV$}
\put(6.7,4.3){$\scriptstyle \sigma^{11}/pb$}
\put(11.9,0.){$ \scriptstyle \sqrt{s}/GeV$}
\put(.5,4.2){$ \scriptstyle (c)$}
\put(8.5,4.2){$ \scriptstyle (d)$}
\end{picture}
\begin{picture}(12,5.9)
\put(-2.5,-9.05){\includegraphics{fig.2e}}
\put(5.5,-9.05){\includegraphics{fig.2f}}
\put(-1.3,4.3){$\scriptstyle \sigma^{11}/pb$}
\put(3.9,0.){$ \scriptstyle \sqrt{s}/GeV$}
\put(6.7,4.3){$\scriptstyle \sigma^{11}/pb$}
\put(11.9,0.){$ \scriptstyle \sqrt{s}/GeV$}
\put(.5,4.2){$ \scriptstyle (e)$}
\put(8.5,4.2){$ \scriptstyle (f)$}
\end{picture}
\caption{\small
        Total cross section $\sigma^{11}$
        for $\mu^+\mu^- \rightarrow \tilde{\chi}^+_1 \tilde{\chi}^-_1$
        in mixed, gaugino and higgsino scenarios 
        with $\mu<0$ ($\mu>0$), 
        a (b), c (d) and e (f) respectively, 
        corresponding to the scenarios
        {A} ({B}), {C} ({D}) and {E} ({F}) 
        of table~\ref{szentab1}.
        In all scenarios $\tan\beta = 5, \ M_A = 350 \ \mbox{GeV},
        \ 
        m_{\tilde{\chi}^+_1 } = 155 \ \mbox{GeV} 
        \ \mbox{and} \  m_{\tilde{\nu}_\mu}=261$~GeV.
        The dashed line corresponds to an energy spread 
	of 150~MeV,
        the solid line to no energy spread.  
\label{cs_scAF}}
\end{figure}

The production cross sections $\sigma^{11}$ (eq.~(\ref{sigmasep}))  
for the scenarios \mbox{A -- F}
are shown in figs.~\ref{cs_scAF} \mbox{a~--~f}.
The heights of the Higgs resonances 
depend both on their total widths and on the Higgs-chargino couplings
(cf.\ eqs.~(\ref{siphi}) and (\ref{propphi}))
\begin{equation}
\sigma_{\phi}^{11}  \propto \vert c^{(\phi)R}_{11}\vert^2 / \Gamma_{\phi}^2, \quad \phi=H,A.
\end{equation} 
The interplay of these parameters (see table \ref{szentab1})
can be observed in fig.~\ref{cs_scAF}.
In our scenarios the pattern of the $A$ resonance is
determined by the width, 
whereas for the $H$ peaks the influence of the different 
$H$-chargino couplings generally predominates.
So the $A$ peaks are of equal height in the mixed and gaugino scenarios 
fig.~\ref{cs_scAF}a and c and larger than in the higgsino scenario
fig.~\ref{cs_scAF}e, inversely proportional to the widths. 
The $H$ resonance is largest in the scenario with the largest
Higgs-chargino coupling fig.~\ref{cs_scAF}a.
Only comparing fig.~\ref{cs_scAF}e and f the relative height of the $H$ peak is
determined by their width   
since the couplings are equal due to an approximate symmetry under 
$\vert\mu\vert\leftrightarrow M_2$.

Essential requirements
for a precise determination of the Higgs-chargino couplings
are distinct resonance peaks and 
a clear separation of the Higgs resonances.
Near threshold
the $A$ resonance peak is suppressed 
by a factor $\beta$, compared to a suppression by
$\beta^3$ of the $H$ resonance.
This effect explains the relative height of the resonances in 
fig.~\ref{cs_scAF}.

Whether the resonances can be separated depends on both the Higgs line
shape and the energy spread of the muon beams. In figs.~\ref{cs_scAF}
\mbox{a -- f} we compare the cross sections without and with a
Gaussian energy spread of 150~MeV which corresponds to an energy
resolution $R \approx 0.06\%$. 

The energy spread clearly suppresses the resonance peaks especially in
scenarios with gaugino-like and higgsino-like light charginos where
the resonances are narrower than in the mixed scenarios.
However, also with an energy spread of 150~MeV the $H$ and $A$ resonances
are
well separated in all scenarios \mbox{(A -- F)}. 

\begin{figure}[ht]
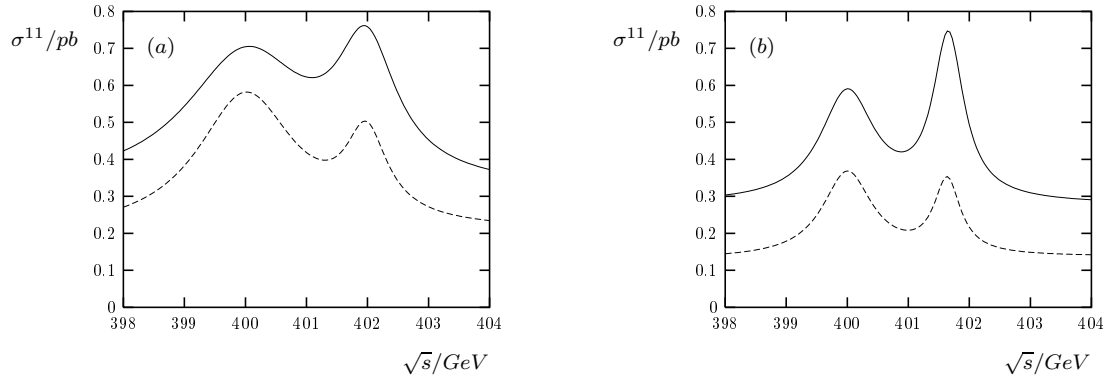

\centering
\setlength{\unitlength}{1cm}
\begin{picture}(12,5.9)
\put(-2.5,-9.05){\includegraphics{fig.3a}}
\put(5.5,-9.05){\includegraphics{fig.3b}}
\put(-1.3,4.3){$\scriptstyle \sigma^{11}/pb$}
\put(3.9,0.){$ \scriptstyle \sqrt{s}/GeV$}
\put(6.7,4.3){$\scriptstyle \sigma^{11}/pb$}
\put(11.9,0.){$ \scriptstyle \sqrt{s}/GeV$}
\put(.5,4.2){$ \scriptstyle (a)$}
\put(8.5,4.2){$ \scriptstyle (b)$}
\end{picture}
\caption{\small
        Total cross section $\sigma^{11}$
        for $\mu^+\mu^- \rightarrow \tilde{\chi}^+_1 \tilde{\chi}^-_1$
        with $\tan\beta =5$, $M_A=400$~GeV,  $m_{\tilde{\nu}_\mu}=261$~GeV and 
        $m_{\tilde{\chi}^+_1 } = 155 \ \mbox{GeV}$ (solid) 
        and
        $m_{\tilde{\chi}^+_1 } = 180 \ \mbox{GeV}$ (dashed).
        (a) shows the mixed scenarios of  table~\ref{szentab2} 
        with $\mu>0$, {B400} and {B180},
        and (b) the gaugino scenarios with $\mu<0$, 
        {C400} and {C180},
        given in table~\ref{szentab2}.
}
\label{sc_ma}
\end{figure}

\begin{figure}[ht]
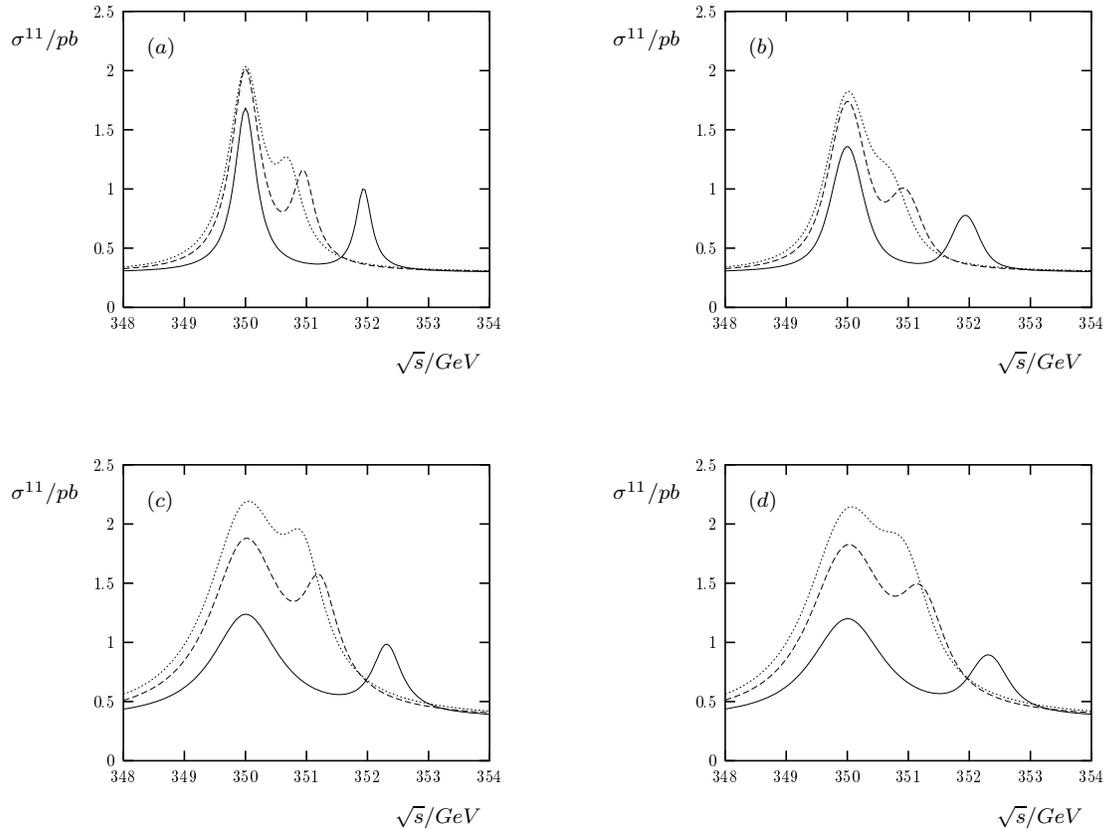

\centering
\setlength{\unitlength}{1cm}
\begin{picture}(12,5.9)
\put(-2.5,-9.05){\includegraphics{fig.4a}}
\put(5.5,-9.05){\includegraphics{fig.4b}}
\put(-1.3,4.3){$\scriptstyle \sigma^{11}/pb$}
\put(3.8,0.){$ \scriptstyle \sqrt{s}/GeV$}
\put(6.7,4.3){$\scriptstyle \sigma^{11}/pb$}
\put(11.8,0.){$ \scriptstyle \sqrt{s}/GeV$}
\put(.5,4.2){$ \scriptstyle (a)$}
\put(8.5,4.2){$ \scriptstyle (b)$}
\end{picture}
\begin{picture}(12,6)
\put(-2.5,-9.05){\includegraphics{fig.4c}}
\put(5.5,-9.05){\includegraphics{fig.4d}}
\put(-1.3,4.3){$\scriptstyle \sigma^{11}/pb$}
\put(3.8,0.){$ \scriptstyle \sqrt{s}/GeV$}
\put(6.7,4.3){$\scriptstyle \sigma^{11}/pb$}
\put(11.8,0.){$ \scriptstyle \sqrt{s}/GeV$}
\put(.5,4.2){$ \scriptstyle (c)$}
\put(8.5,4.2){$ \scriptstyle (d)$}
\end{picture}
\caption{\small
        Dependence on  $\tan\beta$ 
        of the total cross section $\sigma^{11}$
        for $\mu^+\mu^- \rightarrow \tilde{\chi}^+_1 \tilde{\chi}^-_1$
        with $M_A=350$~GeV and $m_{\tilde{\nu}_\mu}=261$~GeV.
        The gaugino scenarios with $\mu<0$, 
        {C}, {C7} and {C8}, 
        are plotted
        without energy spread (a) and with an energy spread of
        150~MeV (b),
        for $\tan\beta =5$ (solid), 7 (dashed) and 8 (dotted),
        and the mixed scenarios with $\mu>0$,
        {B}, {B7} and {B8}, 
        in (c) and (d),
        without and with energy spread respectively and 
        $\tan\beta =5$ (solid), 7 (dashed) and 8 (dotted).
}
\label{sc_tb}
\end{figure}

The influence of the Higgs mass $m_A$ and the chargino mass
$m_{\tilde{\chi}^{\pm}_1}$ is illustrated in fig.~\ref{sc_ma} for
mixed scenarios with $\mu>0$  and for scenarios with a gaugino-like
light chargino and $\mu<0$. In scenarios B400 and C400 with
$m_A=400$~GeV and $m_{\tilde{\chi}^{\pm}_1}=155$~GeV the overlap of
the Higgs resonances is larger than in the corresponding scenarios
with $m_A=350$~GeV and the same chargino mass, see
figs.~\mbox{\ref{cs_scAF} b} and \mbox{\ref{cs_scAF} c}.  The overlap
diminishes when the chargino mass is increased to
$m_{\tilde{\chi}^{\pm}_1}=180$~GeV in scenarios B180 and C180 due to
the smaller phase space of the Higgs decays.

For larger values of $\tan\beta$ the $A$ and $H$ resonances tend to
overlap since the mass difference diminishes. 
As an example 
we compare in fig.~\ref{sc_tb} for $m_A=350$~GeV the total cross
sections for the gaugino scenarios 
{C}, {C7} and {C8} 
with $\tan\beta=5$, $\tan\beta=7$ and
$\tan\beta=8$ respectively, 
without and with an energy spread of 150~MeV.
Without energy spread both resonances are well·
separated up to $\tan\beta=7$ whereas for $\tan\beta=8$ the
$H$ resonance can barely be discerned.·
With energy spread, however, the overlap for $\tan\beta=7$·
is already so large that the resonances
nearly merge.
Here the separation
of the resonance contributions·
may not be possible with a good precision.
The same conclusion applies to other
chargino scenarios, as can be seen for the mixed scenarios 
{B} ($\tan\beta=5$), {B7}·
($\tan\beta=7$) and {B8} ($\tan\beta=8$)
in figs.~\ref{sc_tb}c and fig.~\ref{sc_tb}d without and with
energy spread of 150~MeV, respectively.

\section{Precision measurements of the Higgs-chargino couplings}

The error in the determination of the ratio $x$ of the squared
Higgs-chargino couplings eq.~(\ref{xratio}) depends both on the energy
resolution $R$ of the muon beams and on the error $\Delta
\sigma_B/\sigma_B$ in the measurement of the non-resonant channels
($\gamma$, $Z$, $\tilde{\nu}_{\mu}$ and $h$ exchange as well as
irreducible standard model background) at the $H$ and $A$
resonances.  This background contribution can be estimated from cross
section measurements sufficiently far off the Higgs resonances.

\begin{figure}[ht]
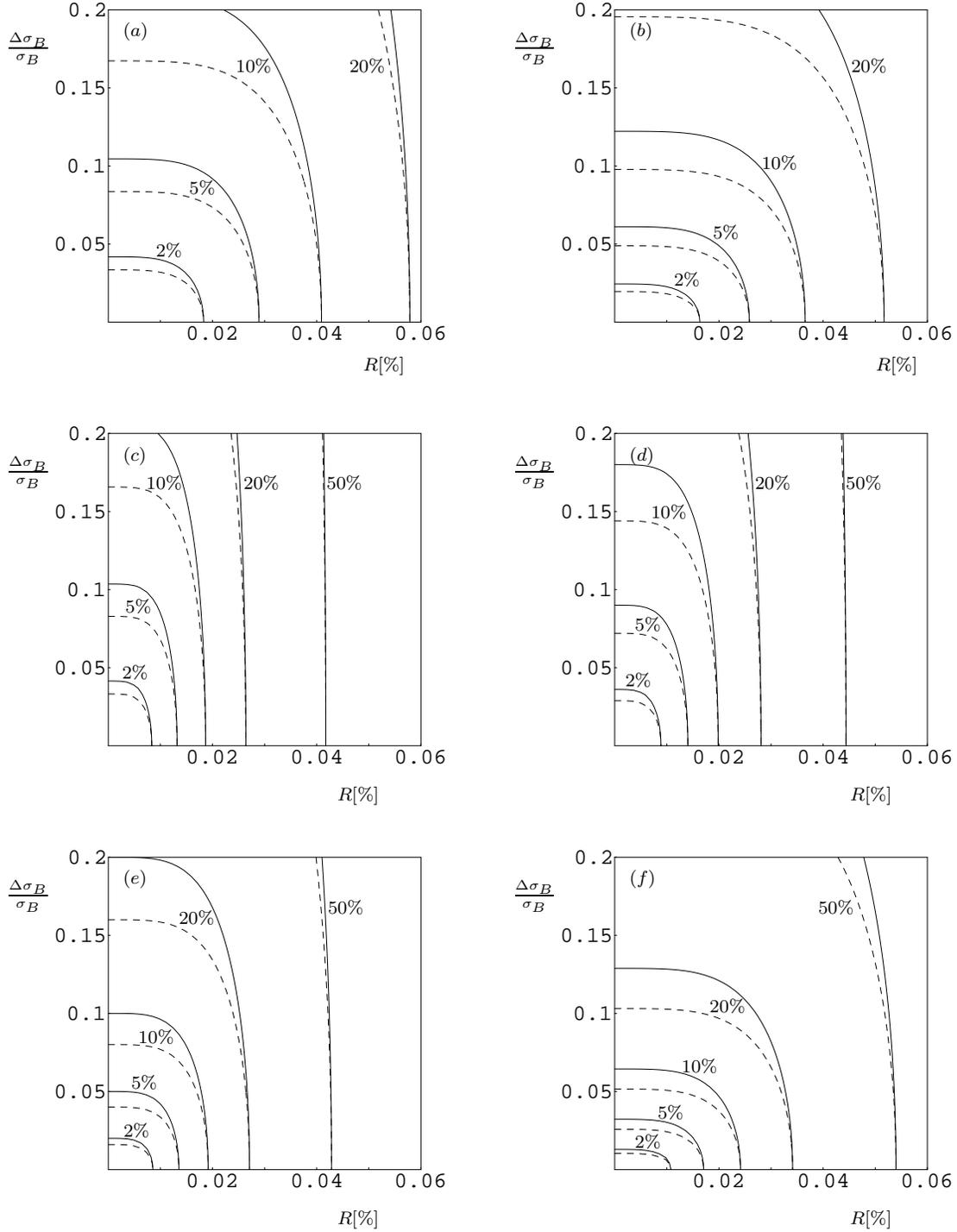

\centering
\setlength{\unitlength}{1cm}
\begin{picture}(13,6.5)
\put(-.3,0.5){\includegraphics{fig.5a}}
\put(7.5,0.5){\includegraphics{fig.5b}}
\put(.8,5.65 ){$ \scriptstyle (a)$}
\put(8.6,5.65){$ \scriptstyle (b)$}
\put(-1.,5.4){$ \scriptstyle 
      \frac{\Delta\sigma_{B}}{
           \sigma_{B}}
$}
\put(6.8,5.4){$ \scriptstyle 
      \frac{\Delta\sigma_{B}}{
           \sigma_{B}}
$}
\put(1.17,2.26){ $\scriptstyle 2 \% $}
\put(1.68,3.23){ $\scriptstyle  5 \% $}
\put(2.4,5.1){ $\scriptstyle  10 \% $}
\put(4.15,5.1){ $\scriptstyle 20 \% $}
\put(9.15,1.8){ $\scriptstyle  2 \% $}
\put(9.75,2.54){ $\scriptstyle  5 \% $}
\put(10.5,3.61){ $\scriptstyle  10 \% $}
\put(11.88,5.1){ $\scriptstyle  20 \% $}
\put(4.5,0.5){$ \scriptstyle R[\%]$}
\put(12.4,0.5){$ \scriptstyle R[\%]$}
\end{picture}
\begin{picture}(13,6.5)
\put(-.3,0.5){\includegraphics{fig.5c}}
\put(7.5,0.5){\includegraphics{fig.5d}}
\put(.8,5.65){$ \scriptstyle (c)$}
\put(8.6,5.65){$ \scriptstyle (d)$}
\put(-1.,5.4){$ \scriptstyle 
      \frac{\Delta\sigma_{B}}{
           \sigma_{B}}
$}
\put(6.8,5.4){$ \scriptstyle 
      \frac{\Delta\sigma_{B}}{
           \sigma_{B}}
$}
\put(.65,2.28){ $\scriptstyle  2 \% $}
\put(.7,3.3){ $\scriptstyle  5 \% $}
\put(1.02,5.2){ $\scriptstyle  10 \% $}
\put(2.52,5.2){ $\scriptstyle  20 \% $}
\put(3.8,5.2){ $\scriptstyle  50 \% $}
\put(8.4,2.17){ $\scriptstyle  2 \% $}
\put(8.55,3.03){ $\scriptstyle  5 \% $}
\put(8.8,4.75){ $\scriptstyle  10 \% $}
\put(10.4,5.2){ $\scriptstyle  20 \% $}
\put(11.8,5.2){ $\scriptstyle  50 \% $}
\put(4.1,0.4){$ \scriptstyle R[\%]$}
\put(12.0,0.4){$ \scriptstyle R[\%]$}
\end{picture}
\begin{picture}(13,6.5)
\put(-.3,0.5){\includegraphics{fig.5e}}
\put(7.5,0.5){\includegraphics{fig.5f}}
\put(.8,5.65){$ \scriptstyle (e)$}
\put(8.6,5.65){$ \scriptstyle (f)$}
\put(-1.,5.4){$ \scriptstyle 
      \frac{\Delta\sigma_{B}}{
           \sigma_{B}}
$}
\put(6.8,5.4){$ \scriptstyle 
      \frac{\Delta\sigma_{B}}{
           \sigma_{B}}
$}
\put(.67,1.75){ $\scriptstyle  2 \% $}
\put(.79,2.5){ $\scriptstyle  5 \% $}
\put(.9,3.2){ $\scriptstyle  10 \% $}
\put(1.52,5.03){ $\scriptstyle  20 \% $}
\put(3.82,5.2){ $\scriptstyle  50 \% $}
\put(8.55,1.58){ $\scriptstyle  2 \% $}
\put(8.9,2.03){ $\scriptstyle  5 \% $}
\put(9.27,2.75){ $\scriptstyle  10 \% $}
\put(9.7,3.67){ $\scriptstyle  20 \% $}
\put(11.37,5.2){ $\scriptstyle  50 \% $}
\put(4.1,0.4){$ \scriptstyle R[\%]$}
\put(12.0,0.4){$ \scriptstyle R[\%]$}
\end{picture}
\caption{
        Relative error in the ratio of the Higgs-chargino couplings $x$
        as a function of the energy resolution and the relative error
        in the non-resonant contributions.
	The irreducible standard model background is neglected (solid)
	and 25\% of the supersymmetric background (dashed).
        Plots \mbox{(a) -- (f)} correspond to the scenarios 
	\mbox{A -- F}
        in table \ref{szentab1}.
}
\label{fig_errors}
\end{figure}

In fig.~\ref{fig_errors} we plot contours of the relative error in the
determination of $x$ in the $R$ and $\Delta \sigma_B/\sigma_B$ plane
for the scenarios A -- F. 
The contours are shown for the two cases
that the irreducible
standard model background is neglected or reduced
to 25 \% of the non-resonant supersymmetric channels by appropriate cuts,
respectively.
For a detailed background analysis Monte Carlo simulations taking into account the detailed detector
characteristics have to be performed and are expected to correspond
to the considered range in
fig.~\ref{fig_errors} \cite {smbackground}.

As a result of the error propagation one
observes a stronger dependence on $R$ than on $\Delta
\sigma_B/\sigma_B$.  Since the energy spread only changes the shape of
the resonance the relative errors in the peak cross sections and in
the widths are correlated.
Generally, an irreducible standard model
background up to 25 \% of the supersymmetric background leads to a slightly
reduced precision for the determination of $x$.

Due to the narrower resonance widths the energy resolution $R$ 
affects the relative error in $x$ in scenarios C, D and E, F
with gaugino-like or higgsino-like light charginos
significantly more than in the mixed
scenarios A and B. The influence of the error in the background
measurement is largest 
in the scenarios with a higgsino-like light
chargino and much smaller in the other chargino mixing scenarios.
In all cases only minor differences appear between the scenarios with positive and negative $\mu$.

In order to achieve a relative error $\Delta x/x < 10\%$ an energy
resolution $R < 0.04\%$ is necessary in the mixed scenarios and less
than $0.02\%$ in the gaugino and higgsino scenarios.  These values lie
in the range between $0.01\%$ and $0.06\%$ of the expected energy
resolution at a muon collider \cite{mucolhiggs,barger1}.  In addition,
the background contributions have to be known with a relative error
$\Delta \sigma_B/\sigma_B<10\%$ in the mixed and gaugino scenarios
whereas in the higgsino scenarios a much higher precision $\Delta
\sigma_B/\sigma_B<6\%$ is necessary.

For a energy resolution $R=0.04\%$ the error in the measurement of
$x$ becomes $\Delta x/x \approx 40\%$ in the scenarios  C and D with
gaugino-like charginos and practically independent of the background
error.  A similar error is expected in scenario E with higgsino-like
charginos, which decreases to 27\% for $\Delta
\sigma_B/\sigma_B<10\%$.

If on the other hand an energy resolution $R=0.01\%$ is achieved and
the contributions of the background channels are well known ($\Delta
\sigma_B/\sigma_B<5\%$ in the mixed and gaugino scenarios and $\Delta
\sigma_B/\sigma_B<2.5\%$ in the higgsino scenarios) the error can be
reduced to the order of a few percent.

\section{Conclusion}

In this paper we have studied chargino pair production at a future
muon collider via resonant heavy Higgs boson exchange in the MSSM.
This process yields large cross sections of up to a few pb in relevant
regions of the supersymmetric parameter space. Due to the sharp energy
resolution that allows to separate the \mbox{CP-even} and
\mbox{CP-odd} resonances a muon collider is an accurate tool to
investigate the Higgs couplings to its decay products. Here we have
focused on the determination of the Higgs-chargino couplings. We have
shown that the ratio of \mbox{$H$-char}gino and \mbox{$A$-char}gino couplings can be
precisely determined independently of the chargino decay mechanism.
This method avoids reference to other experiments and makes only a few
model dependent assumptions, namely the existence of a \mbox{CP-even} and a
\mbox{CP-odd} resonance and the approximate decoupling limit for the
Higgs-muon couplings. 
In representative supersymmetric scenarios we have analyzed the
effect of the energy spread and of the error from the non-resonant channels
including an irreducible standard model background up to 25 \% of the supersymmetric background.
With a good energy resolution a precision
as good as a few percent can be obtained for $\tan \beta < 8$ and $M_A
\leq 400$~GeV, where the Higgs resonances can be separated.

The precision could be further improved by appropriate beam
polarization that enhances the resonant scalar exchange channels and
suppresses the background.  A loss of luminosity
\mbox{\cite{hefreport1,barger1}} as well as effects from initial state
radiation and radiative corrections should be taken into account for
real simulation studies.  The qualitative conclusions of this study,
however, remain unchanged.

\section*{Acknowledgement}

This work was supported by the Bundesministerium f\"ur Bildung und
Forschung, Contract No.~05~7WZ91P (0), by DFG~FR~1064/4-2 and by the
EU~TMR-Network Contract No. HPRN-CT-2000-00149.

\end{document}